\documentclass{article}

\usepackage{epsfig}
\usepackage{amsmath}

\def\lsim{\mathrel{\rlap{\lower4pt\hbox{\hskip1pt$\sim$}}
    \raise1pt\hbox{$<$}}}         
\def\gsim{\mathrel{\rlap{\lower4pt\hbox{\hskip1pt$\sim$}}
    \raise1pt\hbox{$>$}}}         

\def\overleftrightarrow#1{\vbox{\ialign{##\crcr
    $\leftrightarrow$\crcr
    \noalign{\kern 1pt\nointerlineskip}
    $\hfil\displaystyle{#1}\hfil$\crcr}}}

\begin{document}

\hspace{12cm}{\bf TUM/T39-01-01}\\

\begin{center}
{\bf \Large Note on the Impact Parameter Analysis of \\
High Energy Proton Proton Collisions
\footnote{work supported in part by BMBF}}
\end{center}
\begin{center}
{Thorsten Renk$^{a}$} 

{\small \em $^{a}$ Physik Department, Technische Universit\"{a}t M\"{u}nchen, 
D85747 Garching, GERMANY}
\end{center}
\vspace{0.25 in}
\begin{abstract}
Following a prior analysis of measured $pp$ elastic differential
cross sections, the impact parameter representation in terms
of profile functions is calculated from two different parametrizations
of single diffractive dissociation data. The derivative of this quantity
with respect to the collision energy squared $s$ measures the growth rate of the
reaction's blackness. Its distribution in impact parameter space
allows detailed insight into the growth pattern of the total diffractive cross
section and the approaching unitarity limit.
Comparing the results with the elastic case, the different mechanisms
of unitarization of two parametrizations are discussed.

\end{abstract}

\vspace {0.25 in}

\section{Introduction}

It is long known that high energy total hadronic cross
sections grow with  rising center of mass energy $\sqrt{s}$
according to a power law $\left(s/s_0\right)^\epsilon$
and total diffractive cross sections as $(s/s_0)^{2\epsilon}$. Empirically
this behaviour holds for the total single diffractive
cross section up to energies of $\sqrt{s} \sim 30$
GeV, for the total cross section even up to $\sim 1.8$ TeV
and is successfully described
within the framework of Regge theory by the exchange
of various Regge trajectories and the pomeron. However, unitarity
requires that this power law turns over at some point to
agree with the Martin-Froissard bound which demands at most logarithmic growth
$ \sigma \le \sigma_0 \ln^2 \left(\frac{s}{s_0} \right)$. A priori the
point at which this turnover takes place is not determined, and it has been
a vital issue for a long time. 

In several measurements, significant deviations from the power law
given by dominant pomeron exchange
indicating the presence of unitarity limits
have been observed in the case of the total
single diffracitve cross section (e.g. \cite{UA8, UA4, UA4b, ISR, CDF}).

There is, however a quantity which is expected to indicate signals
of the unitarity limit long before they actually show up significantly in the
total cross section. This quantity is the profile function,
an object which is introduced in high energy diffractive reactions
to describe the shape of the collision partners in the plane
transverse to the beam axis. Assuming that the longitudinal momentum
transfer is negligible, it is given by:

\begin{equation}
\label{E-Gamma}
\Gamma({\bf b}) = \frac{1}{2\pi i k} \int d^2 k_t \exp[i {\bf k}_t {\bf b}] 
f({\bf k}_t)
\end{equation}

\noindent
Unitarity constrains the
profile function to satisfy

\begin{equation}
2 \;\text{Re} \;\Gamma({\bf b}) - \Gamma^2({\bf b}) <1,
\end{equation}

\noindent
an expression which reduces to $\Gamma({\bf b}) < 1$
in the limit of vanishing real
part of the scattering amplitude
(the real part of the profile function corresponds
to the imaginary part of the scattering amplitude).

Whereas in the total cross section an average over all impact parameters
is taken, the profile function is directly sensitive to central
collisions in which the unitarity limit is expected to be observed first.
Unfortunately, since the magnitude of the profile function
is strongly influenced by uncertainties in the absolute normalization of
the data, it is not the profile function itself but its
derivative with respect to $s$ which yields the most
interesting observable.

In the following, this property of the profile function will be exploited.
First, the analysis of elastic $pp$ data will be recalled and used
to discuss the requirements necessary to obtain meaningful
results. After that, it is argued that the cross section
$pp \rightarrow pX$ (where X can be any state \emph{excluding the proton}), 
is constrained by unitarity in 
just the same way as the elastic. Due to limitations of data statistics, 
the attention is then focussed on parametrizations, which agree with the
data where available but use two different prescriptions to unitarize
the total cross section. The impact parameter analysis is used to
test these two different prescriptions.

\section{The ISR analysis}

\begin{figure}[!htb]
\begin{center}
\epsfig{file=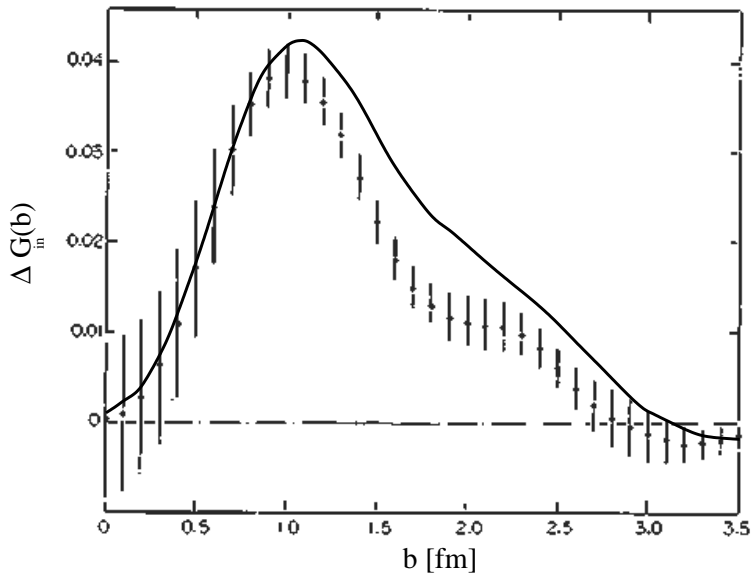, width=10cm}
\end{center}
\caption{\label{F-ISR1}The growth speed of the inelastic overlap integral
$G_{in}({\bf b})$ as function of the
impact parameter as obtained in \cite{Amaldi:1980kd} compared with 
the result of the simplified analysis in which the integration over $|t|$ is
carried over up to $|t| = 1.0$ GeV$^2$ (see text).}
\end{figure}

In \cite{Amaldi:1980kd}, $pp$-elastic scattering data taken at the 
ISR by various
groups have been compiled in order to yield several data
sets of $d\sigma/dt$ for five different $s$. 
The scattering amplitude was now reconstructed assuming
that Im $f({\bf k}_t) \gg$ Re $f({\bf k}_t) \equiv R({\bf k}_t)$ using

\begin{equation}
f({\bf k}_t) = \sqrt{\frac{d \sigma}{d {\bf k}_t} - R^2({\bf k}_t)}
\end{equation}

\noindent
with the small real part taken from dispersion analysis.
The profile function was then calculated using Eq.~(\ref{E-Gamma}).
The growth of the profile with increasing c.m. energy can then be found using

\begin{equation}
\Delta\Gamma({\bf b}) = \frac{d \Gamma({\bf b})}{d \ln s} \bigg|_{s = s_0}
\end{equation}

\noindent
where the derivative are evaluated using averaged
differences of the profiles at each value of ${\bf b}$, therefore
$s_0$ is somewhere between $s = 549$ GeV$^2$ and $s = 3906$ GeV$^2$. 
The actual value of $s_0$ cannot be determined in this way, but it
is not expected that the result depends strongly on $s_0$, it rather
reflects a gross behaviour of the cross section.
In \cite{Amaldi:1980kd}, not the profile function itself is analyzed
in this way, but rather the inelastic overlap integral which is defined
as

\begin{equation}
G_{in}({\bf b}) = 2 \text{Re} \Gamma({\bf b}) - |\Gamma({\bf b})|^2
\end{equation}
which exhibits the same gross features of the growth  
speed at various impact parameters as the profile function.
The result of this
analysis is shown in Fig.~\ref{F-ISR1}.


The most prominent feature is the drop of the blackness 
growth speed at the center $(b=0)$ which is, for very central 
collisions, even compatible 
with zero. This seems to indicate that the corresponding profile functions 
are already approaching the unitarity limit and therefore
cannot grow arbitrarily in the center. The main contribution to the growth
of the total cross section comes from a region of $\sim 1$ fm
which lies at the periphery of the proton.

\begin{figure}[!htb]
\begin{center}
\epsfig{file=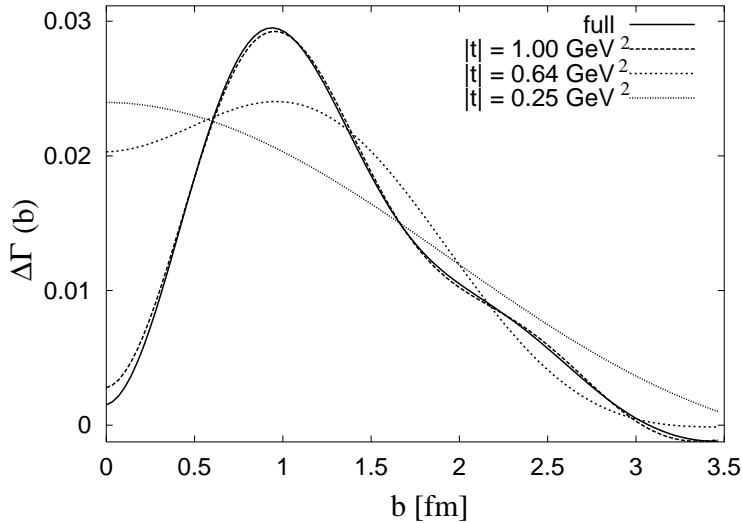, width=10cm}
\end{center}
\caption{\label{F-ISR2}The growth speed of the cross section as function of the
impact parameter for different upper bounds in the Fourier integral
(error bars have been suppressed).}
\end{figure}

The ingredients which are necessary for this analysis are data with:
a) high statistics in $t$ in order to obtain an accurate profile
function, b) different $s$ in order to obtain a reliable derivative
(the low statistics in $s$ is responsible for the large error bars
in Fig.~\ref{F-ISR1}) and c) knowledge of the real part of the scattering
amplitude. In order to test the conditions necessary for the
observation of the central slowdown which is interpreted as a sign for
the unitarity limit, the analysis was redone using the data sets
published in \cite{Tables}, neglecting the real part of the amlitude and 
using variations of the upper bound of the Fourier integral in
Eq.~(\ref{E-Gamma}) in order to test which range in $t$ an
experiment should minimally cover in order to observe this effect. 
The result for the inelastic overlap integral is compared to
the analysis in \cite{Amaldi:1980kd} in Fig.~\ref{F-ISR1} and
the result for the growth speed of the profile function $\Gamma$
is shown in Fig.~\ref{F-ISR2}.


It is obvious that the main signal, namely the drop of $\Delta\Gamma$
for small ${\bf b}$, is still observable in this simplified analysis
in both $\Delta G_{in}({\bf b})$ and $\Delta\Gamma({\bf b})$,
even if we lower the upper bound of the integration down to
0.64 GeV$^2$. This gives confidence that the application of the
same simplified analysis to the case of single diffractive
dissociation may also work and defines the range in $t$ which should
be known experimentally in order to observe this effect
as $|t_{max}|\sim 0.6$ GeV$^2$.

\section{Single diffraction}

Unfortunately, the data situation for the single diffraction process
$pp \rightarrow pX$,
where the final state has a proton and a kinematically separated
hadronic state \emph{excluding the proton}, 
is not satisfactory. Data have been taken
at ISR for several values of $s$ \cite{ISR}; if the dependence on the
mass of the diffractively produced state $X$ is integrated out, only
about 10 data points per set are available to describe $\frac{d\sigma}{dt}$,
starting from $t\sim 0.2$ GeV$^2$. Since the low $t$ region where
the cross section is large gives a dominant contribution in the Fourier 
integral (\ref{E-Gamma}) and the resolution in $t$ is not high,
a sufficiently accurate analysis
based on the measured data alone is not possible. The same is true for
the more recent data obtained by UA4 \cite{UA4, UA4b}, UA8  
\cite{UA8} and CDF \cite{CDF} for vastly different $s$. 
However, existing parameterizations
of the data allow to create 'virtual' data sets. 
In the impact parameter analysis of these virtual data sets, the
unitarization prescription of the parameterizations can be tested.

An adequate description of the shape of the single diffractive differential
cross section is given by Regge theory, although the normalization is suppressed at
high energies relative to the Regge prediction (see \cite{Goulianos:1995wy}). Here
the differential cross section is written as (see e.g. \cite{FF}):

\begin{equation}
\frac{d^2\sigma}{dt dM_X^2}
= \sum_{i,j,k} \frac{\beta_{ik}(0) \beta_{il}(t) \beta_{jl}(t)
g_{ijk}(t)}{16\pi s} \left(\frac{s}{M_X^2}\right)^{\alpha_i(t)+\alpha_j(t)}
(M_X^2)^{\alpha_k(0)},
\end{equation}

\noindent
where $i,j,k$ are all possible combinations of pomeron and other
Regge trajectories, and $\beta_{ik}$ and $g_{ijk}$ the corresponding
vertex functions.
Assuming factorization in the sense that the process can be regarded 
in two steps,
where in the first step the proton emits a pomeron which subsequently,
in the second step, hits the other proton and forms a hadronic state $X$,
the formula can be cast into the form

\begin{equation}
\label{E-Param}
\frac{d^2\sigma}{dt d\xi} = F_{\mathcal{P}/p}(t, \xi) \cdot 
\sigma^{tot}_{\mathcal{P}p}(s')
=[K|F_1(t)|^2\xi^{1-2 \cdot \alpha_{\mathcal{P}}(t)}]
\sigma_0[(\xi s)^{\alpha_{\mathcal{P}}(0)-1}] + \text{Reggeon contributions}.
\end{equation}

\noindent
Here $\xi$ is the momentum fraction that the pomeron carries away from
its parent proton, $s' \equiv M_X^2 \approx \xi s$, $\alpha_{\mathcal{P}}(t)$
is the pomeron trajectory, $F_1(t)$ is the standard Donnachie-Landshoff
form factor \cite{DL} and $K$ is a normalization factor for the pomeron flux. 
Based on this expression, two different parameterizations
have been proposed by Erhan and Schlein \cite{Erhan:2000gs}
and by Goulianos \cite{Goulianos:1999wt}.

It is neither the aim of the present paper to dwell on details
of each parameterization, such as background effects and
possible modifications of the Regge trajectory, nor to compare
their ability to reproduce the data. The focus is rather on a
test of the different prescriptions used for unitarization of
the total cross sections which can be calculated from the two
parameterizations.

In the parameterization by Erhan and Schlein, unitarization
is done via a modification of the pomeron trajectory (which is
usually given as $\alpha_{\mathcal{P}}(t) = \alpha_0 + \alpha' \cdot t$).
Here the trajectory acquires a term which is quadratic in $t$ 
(and unimportant for the purpose of the present analysis since
it affects only the high $t$ range with $t>1$ GeV$^2$
to which this analysis is insensitive)
and the parameters $\alpha_0, \alpha'$ and $\alpha''$ (here $\alpha''$ is the
coefficient of the new quadratic term in the pomeron trajectory)
are modified as a function of $s$ according to

\begin{equation}
\alpha_0(s) = \alpha_0(s_0) + A \cdot \ln \left(
\frac{s}{s_0} \right),\alpha'(s) = \alpha'(s_0) + A' \cdot \ln \left(
\frac{s}{s_0} \right) \text{and} \quad
\alpha''(s) = \alpha''(s_0) + A'' \cdot \ln \left(
\frac{s}{s_0} \right).
\end{equation}

\noindent
For a negative value of $A$ this causes deviations from the power law
and allows to fit the total cross section data.

On the other hand, in the parameterization by Goulianos a
fundamentally different approach is used. Here, 
it is assumed as a working hypothesis that the pomeron flux 
$F_{\mathcal{P}/p}(t,\xi)$ from
the parent proton cannot exceed unity. Therefore the factor
$K$ in Eq.~(\ref{E-Param}) is adjusted in such a way as to meet
this condition. This results in a drastic change in the growth
speed at some critical $s$ beyond which the growth of the
total cross section is slowed down significantly.

It is evident that these two mechanisms to introduce unitarity 
into Eq.~(\ref{E-Param}) are fundamentally different.
Both of them are able to meet the unitarity condition imposed on
the total cross section as seen from the present data. 
The critical question, which now
arises, is the following: Since elastic scattering and 
single diffraction are both constrained by the same unitarity
condition that (assuming negligible real part of the scattering
amplitude) the probability for any interaction must be
smaller than one at some given impact parameter, the same 
behaviour in the growth speed $\Delta\Gamma({\bf b})$ should be seen, namely
a slowing down of central growth as compared to peripheral modes.


\begin{figure}[tb]
\begin{center}
\epsfig{file=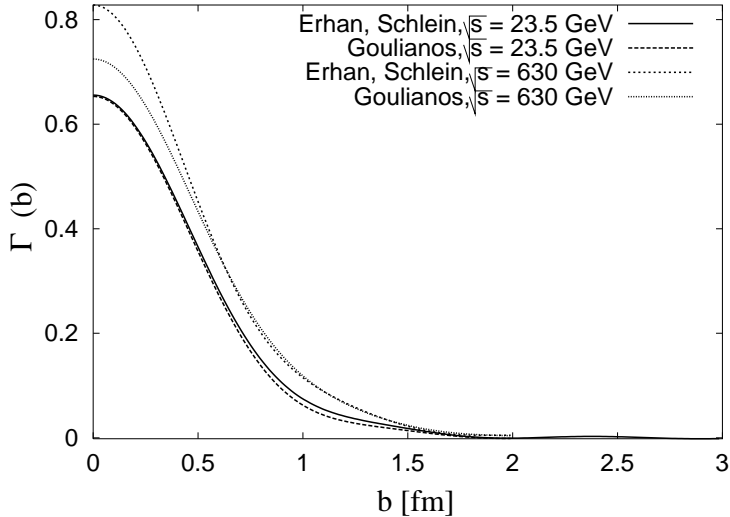, width=10cm}
\end{center}
\caption{\label{F-Profiles}Profile functions obtained by the
parameterizations by Erhan, Schlein and by the one of Goulianos
for different values of $s$.}
\end{figure}



\begin{figure}[!htb]
\begin{center}
\epsfig{file=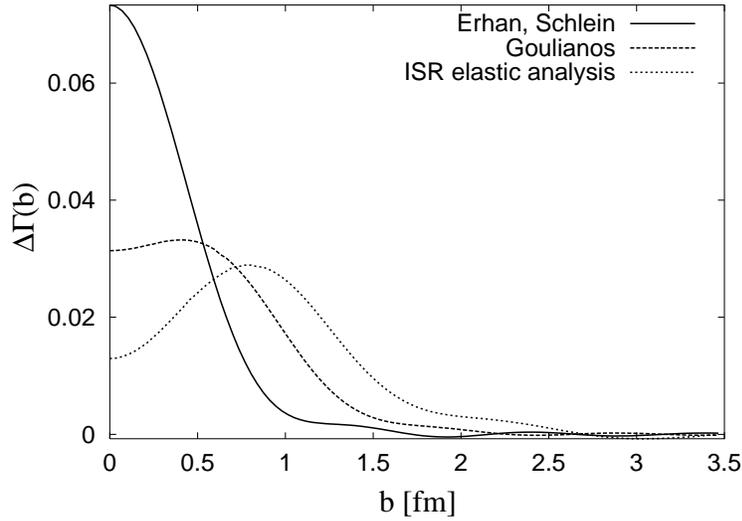, width=10cm}
\end{center}
\caption{\label{F-Result}Impact parameter analysis of the growth speed
$\Delta\Gamma$ based on the parameterization
by Erhan and Schlein and  the one by Goulianos in the ISR
energy range (534 GeV$^2 \le s \le 3906$ GeV$^2$). Shown for comparison
is the result of the analysis of the elastic data as plotted in
Fig.~\ref{F-ISR2}.}
\end{figure}


The actual analysis is done in a way similar to the one done in the
elastic case. Data sets are created from the parameterizations
at values of $s$ in the range of the ISR data, by integrating 
Eq.~\ref{E-Param} over $\xi$  for fixed $t$. The lower limit of
the $\xi$ integration is given by $s'_{min}/s$  where $s'_{min}$
corresponds to the lowest excited state of the proton and the
upper limit is choosen in accordance with the experimental definition
of the published $\frac{d \sigma}{dt}$ data sets as $\xi_{max} = 0.05$.
Both parametrizations have been used exactly as they appear in
\cite{Erhan:2000gs,Goulianos:1999wt}, i.e. including reggeon
contributions.
These data sets are then treated as in the elastic case; the
scattering amplitude is reconstructed assuming a vanishing real part
and the profile function is calculated according to Eq.~(\ref{E-Gamma}). 
Figure~\ref{F-Profiles} shows the resulting profile functions and
Fig.~\ref{F-Result} shows the result of the growth speed analysis.

It is evident from the figure that a unitarization prescription such as the
flux renormalization by Goulianos leads to a picture which is in better
agreement with the one based on the elastic data.
In that (dashed) curve, the dip for central collisions is at least indicated,
whereas the other parameterization (by Erhan and Schlein) 
does not show any sign of
a slower growth of the central blackness --- quite the opposite is
seen, in apparent contradiction to the (dotted) elastic result.

\begin{figure}[!h]
\begin{center}
\epsfig{file=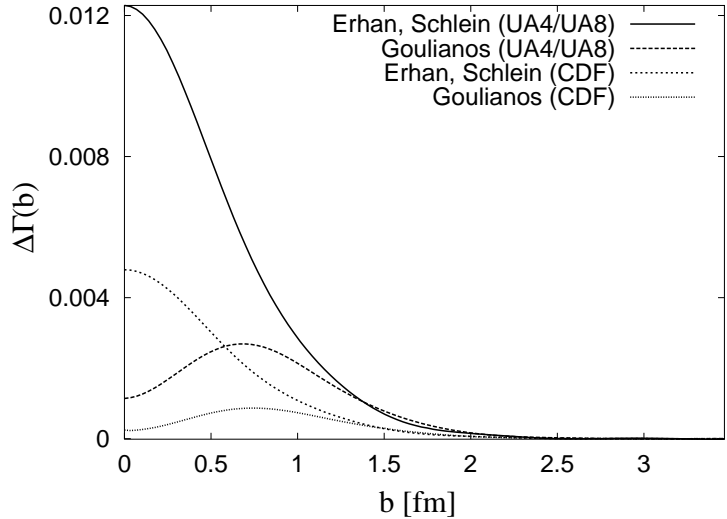, width=10cm}
\end{center}
\caption{\label{F-Result2}Impact parameter analysis of the growth speed
$\Delta\Gamma$ based on the parameterization
by Erhan and Schlein and  the one by Goulianos in the
energy range of the UA4 and UA8 \cite{UA8, UA4, UA4b}
(540 GeV$< \sqrt{s} <$ 630 GeV) and the CDF measurements\cite{CDF} ($\sqrt{s} = 1800$ GeV).}
\end{figure}


The parameterizations can also be tested in the much higher energy range
experimentally accessed by UA4 and UA8. The same analysis as above was made
in the range 540 GeV $< \sqrt{s} <$ 630 GeV and the result is shown in 
Fig.~\ref{F-Result2}. Here, the main features of the ISR energy result
appear again, although somewhat more pronounced.
This can also be seen in Fig.~\ref{F-Profiles}, where the profile
functions of both parameterizations are similar at $s = 550$ GeV,
whereas the one by Erhan and Schlein exhibits a larger central
growth than the one by Goulianos when compared at $\sqrt{s} = 630$ GeV. 
This trend is confirmed looking at the $\sqrt{s} = 1800$ GeV
data obtained by CDF.

Let us conclude this section with a few critial remarks on the
analysis and a summary of basic assumptions. 
First of all, the expression for the profile
function, Eq.~\ref{E-Gamma}, is valid at asymptotic energies only.
At finite energies, a mathematically correct treatment has been developed
(see e.g.\cite{FE}) and, in principle, 
should be used. Next, throughout the analysis,
the real part of the scattering amplitude has been neglected.
This appears to be justified by looking at Fig.~\ref{F-ISR1},
where this approximation amounts only to a small difference for ISR
energies. Similar assumptions are used in the standard
analysis of high-energy elastic hadron scattering.
The last issue concerns the form of the parametrizations used.
Here, only one of the diagrams shown by Ross and Yam \cite{RY}
has been used in both cases discussed in this paper. 
On the other hand, both parametrizations
are able to account for the data where available. Clearly
the shape of the profile functions resulting from the
present analysis is central, in spite of the peripheral nature
of diffraction. However, the conclusions concerning
the unitarization of the parametrizations do not
depend on the last two issues.

\section{Summary}

We have shown that the impact parameter analysis is a useful tool
to investigate how effects caused by unitarity limits are
distributed across the impact parameter space. The strongest
manifestations of such effects should be found for central collision where
the blackness is largest, resulting in a slowing down of the growth
speed of the profile for small impact parameters. 
This method can not only be used for measured data but also
in order to analyse unitarization prescriptions in empirical
parameterizations. The reaction $pp\rightarrow pX$
has been considered here to demonstrate that
unitarization by flux renormalization is closer
to what one would expect, guided by the analysis of the
elastic scattering data, than unitarization by introduction
of an $s$-dependent pomeron intercept.

I thank G. Piller and W. Weise for helpful comments and discussions.


\begin{thebibliography}{99}

\bibitem{UA8}
A.~Brandt et al.  (UA8 Collaboration),
Nucl.\ Phys.\  {\bf B514}, (1998) 3 


\bibitem{UA4}
M. Bozzo et al. (UA4 Collaboration),
Phys. Lett. {\bf B 136} (1984) 217

\bibitem{UA4b}
M. Bernard et al. (UA4 Collaboration),
Phys. Lett. {\bf B 186} (1987) 227


\bibitem{ISR}
M. G. Albrow et al., Nucl. Phys. {\bf B 54} (1973) 6;
M. G. Albrow et al., Nucl. Phys. {\bf B 72} (1974) 376

\bibitem{CDF}
F.~Abe et al. (CDF Collaboration),
Phys. Rev. {\bf D50} (1994) 5535


\bibitem{Amaldi:1980kd}
U.~Amaldi and K.~R.~Schubert,
Nucl.\ Phys.\  {\bf B166} (1980) 301


\bibitem{Tables}
K. R. Schubert, \emph{Tables on nucleon nucleon scattering}, 
Landolt-B\"{o}rnstein, 1/9A (1979) 

\bibitem{Goulianos:1995wy}
K.~Goulianos,
Phys.\ Lett.\ B {\bf 358} (1995) 379


\bibitem{FF}
R. D. Field and G. C. Fox, Nucl. Phys. {\bf B 80} (1974) 367


\bibitem{DL}
A. Donnachie and P. V. Landshoff, Nucl. Phys. {\bf B 231} (1984) 189;
A. Donnachie and P. V. Landshoff, Nucl. Phys. {\bf B 267} (1986) 690


\bibitem{Erhan:2000gs}
S.~Erhan and P.~E.~Schlein,
Phys.\ Lett.\  {\bf B481} (2000) 177

\bibitem{Goulianos:1999wt}
K.~Goulianos and J.~Montanha,
Phys.\ Rev.\  {\bf D59} (1999) 114017

\bibitem{FE}
T.~Adachi and T.~Kotani,
Progr.~Theor.~Phys.~Suppl., Extra Number (1965) 316;
Progr.~Theor.~Phys. {\bf 35} (1966) 463; {\bf 36} (1966) 745; 
{\bf 37-38} (1966) 297

\bibitem{RY}
M.~Ross and Y.~Y.~Yam,
Phys. Rev. Lett. 19 (1967) 546

\end{thebibliography}
\end{document}